\documentstyle[aaspp4,11pt]{article}
\tighten

\begin{document}

\title{The Effect of Expansion on Mass Entrainment and \\ Stability of
Super-Alfv\'enic Jets}

\author{Alexander Rosen and Philip E.\ Hardee}

\affil{Department of Physics \& Astronomy \\ University of Alabama \\ Tuscaloosa, AL 35487 \\ rosen@eclipse.astr.ua.edu, hardee@athena.astr.ua.edu}

\begin{abstract}

We extend investigations of mass entrainment by jets, which previously
have focused on cylindrical supermagnetosonic jets and expanding
trans-Alfv\'enic jets, to a set of expanding supermagnetosonic jets.
We precess these jets at the origin to excite the helical mode of the
Kelvin-Helmholtz (or KH) instability, in order to compare the results
with predictions from linear stability analysis.  We analyze this
simulation set for the spatial development of magnetized mass, which we
interpret as jet plus entrained, initially unmagnetized external
mass.  As with the previous simulation sets, we find that the growth
of magnetized mass is associated with the growth of the KH instability
through linear, nonlinear, and saturated stages and with the expansion
of magnetized material in simulated observations of the jet.  From
comparison of measured wavelengths and wave speeds with the predictions
from linear stability analysis, we see evidence that the KH instability
is the primary cause for mass entrainment in these simulations, and
that the expansion reduces the rate of mass entrainment.  This reduced
rate can be observed as a somewhat greater distance between the two
transition points separating the three stages of expansion.

\end{abstract}


\lefthead{ROSEN \& HARDEE}
\righthead{Expanding Supermagnetosonic Jets}

\section{INTRODUCTION}

Mass entrainment of an external medium into a jet flow affects the
spatial development of the jet and may be a crucial factor in the
division between FR I and FR II (Fanaroff \& Riley 1974) radio
sources.  For example, deceleration, presumably via mass entrainment,
of a turbulent, supersonic jet is one of the basic assumptions in the
model of Bicknell (1994, 1996) that reproduces the break between FR I
and FR II sources in the radio-optical plane.  Additionally, there is
evidence for jet slowing in the low-power radio source B2 1144+35,
which is observed to decelerate from 0.95$c$ to 0.02$c$ between 20 pc
and 24 kpc from the core (Giovannini et al.\ 1999).  Mass entrainment
may be important even in slower jet flows, such as the much slower
(0.1$c$) parsec-scale jets in Seyfert galaxies (Ulvestad et
al.\ 1999).  Since our simulations do not include radiative cooling,
which is a crucial element in the energetics of protostellar jets, the
simulations we present here are not strictly applicable to those
systems, although perhaps they are appropriate for the jets in Galactic
superluminals.

Rosen et al.\ (1999, hereafter RHCJ) confirmed that a reasonably strong
magnetic field can slow mass entrainment by a three-dimensional,
nonrelativistic, cylindrical supermagnetosonic jet.  RHCJ showed that
the spatial growth of mass entrainment is a three-stage process that
corresponds to the growth of the Kelvin-Helmholtz (KH) instability,
from a linear stage through a nonlinear stage to a stage where the
growth has become saturated.  RHCJ also found three stages of growth in
the width of simulated radio intensity maps of the jets, and that these
three stages in the width are associated with those of the growth of
the jet plus entrained mass.  However, some detailed differences did
exist in the position of the transition between stages as a function of
jet density.  These differences usually occurred for jets lighter than
the external medium, and took the form of continued mass entrainment
even though the width in a simulated intensity image of the 
jet remained constant.  A jet that entrains mass could have
an observable spine-sheath morphology, in which the spine is more
disrupted in a jet that is lighter than the external medium.  In fact,
recent application of the ``spectral tomography" process (Katz-Stone et
al.\ 1999) has revealed the presence of a low intensity, steep spectrum
sheath surrounding a high intensity, flat spectrum core or jet spine in
a few wide-angle tail radio sources.  In addition, RHCJ found that the
linear growth of jet plus entrained mass was related to a linear progression
as higher order, small amplitude KH modes with short growth lengths are overwhelmed by lower order, larger amplitude KH modes with larger
growth lengths.

In this paper, we extend our mass entrainment analysis to conical
supermagnetosonic jets that have inlet conditions similar to some of
the jets in RHCJ.  Based on the KH linear stability analysis for
supermagnetosonic jets in Hardee, Clarke, \& Rosen (1997, hereafter
HCR), the growth length at maximum growth (i.e., the inverse of the
spatial growth rate at maximum growth) is proportional to the radius of
the jet.  Additionally, the maximum growth rates are reduced along an
adiabatically cooling jet, in which the Mach number increases with the
jet expansion (Hardee 1986).  Thus, in general on supermagnetosonic jets the
instability should grow more slowly as the radius expands than on a
cylindrical jet.  We investigate whether expanding supermagnetosonic
jets are more stable than similar cylindrical jets and also whether
expanding jets show the similar association between the spatial
development of mass entrainment, the spatial development of the KH
instability and the spatial development of the apparent width of
magnetized material in simulated radio intensity images that has been
seen previously in cylindrical jet simulations (RHCJ).

Examination of the potentially smaller effects of the KH instability on
conical jets is an important extension of previous work, as there are
observational examples of conical jets, some of which expand by orders
of magnitude and remain undisrupted for more than 100 kpc.  For
example, the jet in M87 expands at a nearly constant rate in a recent
high resolution (0$\farcs 1\approx$ 8.2 pc) radio image of the source
out to knot A (Biretta, Zhou, \& Owen 1995).  Another example of jet
expansion is provided by NGC 6251, which has a jet that passes through
a series of expansions, each of which has an average half-opening angle
of $\approx$4\arcdeg\ for at least 10 kiloparsecs in length (Perley et
al.\ 1984).  The jets in both of these radio sources appear more
laminar than turbulent in the regions where the jets are conical.

Since we aim to investigate the spatial growth of the KH instability
and mass entrainment, the simulations we discuss here span a
significant axial length and are considered completed only after much
of the computed region reaches a quasi-steady state.  This approach
contrasts with that of numerical simulations with periodic boundary
conditions along the jet axis, e.g., Bassett \& Woodward (1995), Bodo
et al.\ (1995), and Bodo et al.\ (1998), which focus on temporal
growth.  This paper is organized as follows: in \S 2, we explain the
initialization of the 3D MHD simulations, and in \S 3, we discuss, in
order, the growth of mass entrainment, the observational
characteristics associated with mass entrainment in simulated (radio
wavelength) intensity and polarization images, and the testing of the
presence and measurable consequences of the KH instability.

\section{INITIALIZATION OF NUMERICAL SIMULATIONS}

These simulations were performed with the three-dimensional
magnetohydrodynamics code ZEUS-3D, an Eulerian finite-difference code
with the Consistent Method of Characteristics (CMoC) that solves the
transverse momentum transport and magnetic induction equations
simultaneously and in a planar split fashion (Clarke 1996).
Interpolations are carried out by a second-order accurate monotonic
upwinded time-centered scheme (van Leer 1977) and a von
Neumann-Richtmyer artificial viscosity is used to stabilize shocks.
The code has been thoroughly tested via MHD test suites as described by
Stone et al.\ (1992) and Clarke (1996) to establish the reliability of
the techniques.

The useful computational domain for the three expanding jet simulations
discussed here is a 3D Cartesian grid of 325 $\times$ 130 $\times$ 130
zones, which required roughly 105 Mwords of memory on the Cray C90 at
the Pittsburgh Supercomputer Center (or PSC).  The central thirty zones
span the initial jet diameter, 2$R_0$, along the transverse Cartesian
axes (the $x$-axis and $y$-axis). Outside the uniform grid zones, the
grid zones are ratioed where each subsequent zone increases in size by
a factor of 1.050.  The 130 transverse zones span a total distance of
30$R_0$.  Along the jet or $z$-axis, 225 uniform zones span the
interval 0 $< z/R_0 <$ 30 and an additional 100 ratioed zones span an
additional 30$R_0$ with each subsequent zone increasing in size by a
factor of 1.015.  Outflow boundary conditions are used except where the
jet enters the grid, where inflow boundary conditions are used.

The three simulations are initialized at the inlet with parameter
values similar to simulations in HCR and RHCJ.  Specifically,
simulation A here is equivalent to run C in RHCJ, simulation B is
equivalent to run A, and simulation C is equivalent to run E.
Simulations A and B contain jets that are denser than the external
medium and simulation C is of a light jet.  Details of jet properties
are listed in Table \ref{para}.  While some simulations in RHCJ had a
primarily toroidal magnetic field, all of the simulations here have a
primarily axial magnetic field.  Limiting this study to axial field
jets was done to maximize the instability of the jet, since we
anticipate expanding jets to be more stable and we know that a toroidal
magnetic field can have an additional stabilizing effect (compared with
a primarily axial magnetic field of the same strength, e.g., Appl \&
Camenzind 1992; RHCJ).  A smaller jet velocity, $u$, is used in order
to reduce all of the Mach numbers (i.e., sonic, Alfv\'enic, and
magnetosonic) because we wish the instability to grow significantly
while the jet is within the computational domain and the growth length
of the instability is roughly proportional to the product of jet radius
and magnetosonic Mach number, with $M_{ms} \equiv u/a_{ms}$.  The
magnetosonic speed is defined as $a_{ms}^2 \equiv a_{jt}^2 + V_A^2$,
where the internal jet sound speed and the Alfv\'en speed are defined
by $a_{jt}^2 \equiv \Gamma p/\rho$ and $V_A^2 \equiv B^2/4\pi\rho$,
respectively, and we assume the adiabatic gas law, $P \propto \rho^\Gamma$
with $\Gamma$ = 5/3.  We have analyzed the simulations at a somewhat later
time, at $\tau$ = 45 dynamical times for simulations A and B compared
with $\tau$ = 36 in RHCJ and $\tau$ = 16 for simulation C compared with
$\tau$ = 14 in RHCJ, where $\tau \equiv t a_{ex}/R_0$ and $a_{ex}$ is
the external sound speed.  The jets in all of the expanding jet
simulations pass through at least 3 flow-through times at $z$ =
50$R_0$.  Each simulation required about 40 CPU-hours on the Cray C90
at PSC.

The jet is initialized across the computational domain with a uniform
density, $\rho_{jt}$, and initial radius, $R_0$.  The magnetic field
within the jet is initialized with a uniform axial component, $B_z$,
for $r < R_0$, where $r$ is the radial position, and a toroidal
magnetic component, $B_\phi$, with a radial profile that has the form
$B_\phi(r) = B_\phi^{pk} \sin^2[\pi f(r)]$, where $B_\phi^{pk}$ is the
maximum value for $B_\phi$ and for $r<r_{pk}$, $f(r)=0.5(r/r_{pk})^a$,
and for $r_{pk}<r<r_{\max }$, $ f(r)=1.0-0.5\left[ \left( 1-r/r_{\max
}\right) /(1-r_{pk}/r_{\max })\right] ^b$.  In these simulations $a$ =
1.106 and $b$ = 0.885, which leads to a FWHM of the $B_\phi$ radial profile
equal to 0.5R$_0$.  This radial profile for $B_\phi$, which
sets $B_\phi$ to zero at $r/R_0$ = 0.0 and for $r/R_0 \ge$ 0.9, is not
physically motivated, but is mathematically well-behaved.  The magnetic
field is set to zero in the external medium.  The jet thermal pressure
is initially modified to satisfy an equation of hydromagnetic
equilibrium,

\begin{equation} {d \over {dr}}\left (p_{jt}(r) + {B_z^2(r) \over {8
\pi}} + {B_\phi^2(r) \over {8 \pi}}\right) = - {B_\phi^2(r) \over {4
\pi}}, \end{equation}

\noindent where the term on the right hand side describes the effects
of magnetic tension.  Note that this equilibrium condition is one of
static balance, but does not necessarily lead to hydromagnetic dynamic
equilibrium, which is determined by the condition $\nabla \times ({\bf
u}\times {\bf B})=0$.  Owing to the small toroidal field in each of
these simulations, this dynamic equilibrium is nearly satisfied and we
have verified that no significant azimuthal velocity occurs within the
jet near the inlet.  The inlet values for $B_z$ and $B_\phi^{pk}$ are
given in footnotes to Table \ref{para} in the units used by ZEUS-3D,
which sets the permeability of free space to 1.  While the thermal
pressure has been modified to satisfy the radial static equilibrium,
the departure from a ``top hat" pressure distribution is small.  The
typical plasma $\beta_0 \equiv$ 8$\pi p_{jt}/B_{z}^2$ is 200 in 
simulation A and 1.1 in simulations B and C.

In the simulations the external medium is isothermal and the external
density, $\rho _{ex}(z)$, declines to produce a pressure gradient that
is designed to lead to a constant expansion,
$R_{jt}(z)=(1+z/120R_0)R_0$, of a constant velocity adiabatic jet
containing a uniform axial magnetic field and an internal toroidal
magnetic field that provides some confinement, i.e.,

\begin{equation} \rho_{ex}(z) =  {\left (\varrho^{-10/3} +
C_1\varrho^{-4} - C_2 \varrho^{-2}\right )  \over C_3} \end{equation}

\noindent where $\varrho = R(z)/R_0$, $R(z)$ is the jet radius at axial
position $z$, the constants $C_1$ and $C_2$ are based on the ratios at
the inlet of jet thermal pressure to magnetic pressure from the axial
component (in $C_1$) or from the toroidal component (in $C_2$), and
$C_3$ is a normalization constant equal to  1 + $C_1$ - $C_2$.  We have
confirmed that the initial expansion occurs quickly, within 5--7
dynamical times for the dense jet simulations and 1--2.5 dynamical
times for the light jet simulation.

A difference between these simulations and their cylindrical
equivalents in RHCJ is a slightly reduced angular precessional
frequency, $\omega$ (see Table \ref{prec}).  We decrease $\omega$ to
obtain the same normalized precession frequency, $\omega R(z)/a_{ex}$,
at some small (but nontrivial) distance from the inlet in each
expanding jet as in the appropriate cylindrical equivalent.  The
spatial KH stability analysis of expanding jets suggests that at peak
growth the normalized precession frequency is constant (Hardee 1986);
our choice for the precession frequency, $\omega$, is below that for
peak growth, but $\omega R(z)/a_{ex}$ will increase with jet
expansion.  This avoids the unresponsiveness of a jet to too high a
precessional frequency.  The precession is in the counterclockwise
sense at the inlet and should yield a clockwise spatial helical twist
when viewed downstream from the origin/inlet.  The simulations are
analyzed at 6.5, 7.9, and 3.6 precessional periods for simulations A,
B, and C, respectively.  Another difference between the simulation sets
is that the radial variation of the toroidal magnetic field component,
$B_\phi$, peaks at 0.5$R_0$ in the expanding jet simulations, while in
all three cylindrically equivalent simulations in RHCJ this maximum in
$B_\phi$ occurred at $r$ = 0.8$R_0$.  However, since the toroidal field
is not dynamically important in these simulations, this difference is
inconsequential.

\section{RESULTS}

\subsection{Linear Mass Density as a Mass Entrainment Estimator}

Since the code does not include a tracer of jet material, we use
magnetic field and axial velocity criteria to estimate the mass
associated with the jet plus any external medium that has been
accelerated or mixed with the jet sufficiently to be called
``entrained" (RHCJ).  Such a velocity criterion has been used previously 
as an estimate for mass entrainment (Bassett \& Woodward 1995, Loken
1996).  Since RHCJ found that the choice of the magnetic field or axial
velocity criteria result in similar values for the estimated jet plus
entrained mass, we have chosen to use the magnetic field criterion
alone in this paper.  As in RHCJ, we define the linear mass density,
$\sigma$, at any point along the jet as

\begin{equation} \sigma(z) = \int_{\rm A} f\rho~dx\,dy,
\end{equation} 

\noindent where A is the cross-sectional area of the computational
domain at $z$, and $f$ is a switch set to 1 if the local magnetic field
is above a threshold value and $f$ = 0 otherwise.  We set the threshold
to $0.04 B_{max}$, where $B_{max}^2 \equiv B_{\phi}^{pk}(z)^2 +
B_{z}(z)^2$.  In a steady expanding flow, $B_\phi$ varies as
$R(z)^{-1}$ and $B_z$ varies as $R(z)^{-2}$.  
While the choice of the 4\% level is somewhat arbitrary,
we believe that a threshold at this value is a useful demarkation
between ``mixed" (i.e., jet plus entrained material) and ``unmixed"
media.  The unmixed external medium may contain some small magnetic 
field from numerical diffusion and should not be considered as
entrained.  In our previous simulations, this 4\% $B_{max}$ level
corresponded to external material accelerated up to $\sim$ 15\% of the
jet speed in the dense jets and $\sim$ 4\% of the jet speed in the
light jet.  In a smooth flow of constant jet speed without entrainment,
the linear mass density remains constant for both cylindrical and
expanding jets.  In what follows, $\sigma(z)$ is the linear mass
density for the axial position $z$ normalized to the initial jet linear
mass density.

We display the spatial development of $\sigma(z)$ for the expanding jet
simulations and for the equivalent cylindrical jet simulations in
Figure \ref{sig}.  In addition, the bottom panel of Figure \ref{sig}
shows the average axial velocity of the magnetized (with $B >
0.04B_{max}$) material, weighted by the linear mass density
(i.e.\ $\langle v_z\rangle = \int f\rho v_z~dx\,dy/\int f\rho~dx\,dy$),
along $z$.  From a comparison of the linear mass density in equivalent
cylindrical and conical jet simulations, the jet plus entrained mass at
the $z =60R_0$ boundary is reduced by roughly one-half in the dense
jets and by at least two-thirds in the light jet (see Table
\ref{tran}).  Calculations of the mean density of entrained material
(see Figure \ref{aved}), which is defined as the mixed linear mass
density less the initial jet linear mass density and corrected for
different cross-sectional areas, shows that this mean density is
similar to the initial external medium density.  Thus, the primary
reason for the reduction in entrained mass is the reduced external
density in the expanding jet simulations, which is required by total
pressure balance between the expanding jet and the isothermal external
medium.  For the parameters that we have simulated, the reduction of
$\sigma$ in the expanding jets appears similar to (for the dense jets)
or greater than (for the light jet) the reduction caused by changing
the magnetic field configuration from primarily axial to primarily
toroidal in the cylindrical jet simulations of RHCJ.

In RHCJ, the linear mass density plots showed regions with different
rates of increase of $\sigma(z)$ with $z$, which we interpreted as
different mass entrainment rates.  Specifically, successive regions of
slow, fast, and no mass entrainment (i.e., mass mixed with the fast
moving jet as measured by our magnetic criterion) were found, and there
was evidence that the three stages of growth were associated with
linear, nonlinear and saturated growth of the KH instability.  In both
dense jets (simulations A and B), the rate of increase of linear mass
density with axial position remains approximately constant for the
entire range plotted.  However, for $z/R_0 \gtrsim 20$ the fluctuation
of $\sigma$ about this constant rate also increases with $z$, which is
consistent with the notion of increasingly large ``gulping" vortices,
such as those discussed by DeYoung (1996), at the jet surface.  This in
turn is a signature of a developing KH instability.  In addition, there
are noticeable changes in $\partial \sigma/\partial z$ near this
position ($z/R_0 \sim$ 20) when the linear mass density is computed
with a smaller threshold value (i.e., smaller than 4\%).  If the onset
of the fluctuations in $\sigma$ in the dense expanding jet simulations
is associated with the first transition point, then the first
transition points of the dense jet simulations occur at roughly the
same axial positions as that of the first transition points in their
cylindrical equivalents (see Table \ref{tran}).

Of the dense expanding and cylindrical jet simulations, only the
linear mass density in the cylindrical equivalent to the weak
field jet (simulation A) could be interpreted as leveling off
before the outer $z$ boundary.  From our analysis of the growth of
magnetized mass in the simulations here and in RHCJ, we suspect 
that the linear mass density in the dense expanding jets would 
eventually saturate beyond the second transition (saturation) 
position in the cylindrical jets.  If this is true, then the distance
between the first and second transition points would be larger
in the dense expanding jets than in their cylindrical equivalents.  

In the expanding light jet (simulation C), there are three regions of
differing rates of increase of $\sigma(z)$ with $z$.  Estimated rates
of increase are shown for the second (fast growth) stage in the light
jet simulations in Figure \ref{sig}.  Transition points between each
region are listed in Table \ref{tran} for the expanding and cylindrical
jets simulations.  The first transition point is significantly closer
to the inlet in the expanding jet simulation than in the cylindrical
jet simulation.  Since the second transition point is at roughly the
same position in both the expanding and cylindrical jets, we see that
the distance between the first and second transition points is larger
in the expanding jet than in the cylindrical jet.  As with the dense
jets, the light expanding jet has a smaller spatial rate of increase of
$\sigma(z)$ with $z$ in the fast growth region than the light cylindrical
jet.

We postulate that the KH instability is associated with the mass
entrainment in these expanding jet simulations and compare the
transition positions in Table \ref{tran} with estimates of the growth
lengths of the KH instability determined from a linear analysis.
Estimated growth lengths of the helical surface wave ($n$ = 1, $m$ = 0)
for both the expanding and cylindrical simulations are listed in Table
\ref{grow}.  All of these estimates are from the simple approximation
for the growth length at the maximum growth rate, $l^* = k_I^{*\,-1} =  -
(2M^{ms}_{jt}R)/\ln(4\omega_{nm}R/a_{ex})$ (equation 5 in HCR), where
$\omega_{nm}R/a_{ex} = (n + 2m + 1/2)\pi/2$ (equation 4a in HCR).  Note
that this method consistently underestimated the values found from more
sophisticated root-finding techniques for the cylindrical jet
simulations in RHCJ.  Since the growth length depends strongly on the
jet radius, for the expanding jet numerical integration was used to
compute the number of effective e-folding lengths, $N_e$, such that
$N_e = \int^z_{z=0} k_I(z) dz$ (equation [17] in Hardee 1986).  From
the smaller inlet magnetosonic Mach number, the expanding jets have a
slightly smaller initial e-folding length than their cylindrical 
equivalents.  The similar e-folding length for the expanding 
and cylindrical jets is consistent with the similar (or smaller in the
light expanding jet) axial position of the first transition point.  The
increasing radius of an expanding jet leads to a larger distance
for 5 e-foldings, which suggests that the expanding jet simulations
should remain more stable farther down the grid.  This greater
stability is consistent with the larger distance spanned between first
and second transition points in the expanding jets than in their
cylindrical equivalents.

The average axial velocity of the magnetized material decreases at the
same rate with $z$ in the dense expanding jet simulations (simulations
A and B).  This is similar to the constant rate of increase in the
linear mass density, although the significant fluctuations in $\sigma$
are absent in $\langle v_z\rangle$.  In the light expanding jet
(simulation C), there are at least three regions where the average
axial velocity behaves differently, and these regions are roughly
coincident with the regions of differing mass entrainment rates.
Specifically, there is an initial region out to $z/R_0 \lesssim$ 14
where there are very small fluctuations in the average axial velocity,
which changes its smooth rate of decline at $z/R_0 \approx$ 8.  Beyond
this initial region with small fluctuations, there are significant
fluctuations superimposed on a roughly constant decrease in the average
axial velocity for 14 $\lesssim z/R_0 \lesssim$ 35, and there are small
fluctuations about an approximately constant $\langle v_z\rangle$
thereafter.  The greater reduction in average velocity in the light jet
simulation compared with the dense jet simulations is a result of
mixing of the jet with the denser external medium.  The region of
approximately constant $\langle v_z\rangle$ in the light expanding jet
suggests that any mixing has ceased and is consistent with the
saturation of the KH instability.

We calculated (but do not show) the transverse area covered by
magnetized material with $B > 0.04 B_{max}$.  In general, the area
occupied by magnetized material is approximately the same for the dense
expanding jets and for their cylindrical equivalents at any computed
$z$.  The area occupied by magnetized material in the light expanding
jet simulation is slightly smaller than the area of its cylindrical
equivalent.  Therefore, the relative increase in the area occupied by
magnetized material, i.e.\ (magnetized area)/(initial jet area), is
smaller in all expanding jet simulations than in their cylindrical
equivalents.  This is consistent with a smaller spatial growth rate of
the KH instability.  In addition, we computed (but do not show) the
linear momentum flux density, $\sigma v_z^2$, of the mixed and unmixed
regions.  As in the trans-Alfv\'enic jets studied in Hardee \& Rosen
(1999), the mixed material carries the bulk of the momentum flux.  The
maximum momentum flux density carried by the unmixed material in any
computed plane of transverse zones is about 10\% of the initial jet
momentum flux density for the dense jets (simulations A and B) and
about 5\% for the light jet (simulation C).

\subsection{Comparison of Linear Mass Density and Simulated Radio Images}

In Figure \ref{radio} we display maps of simulated total radio
intensity, which are line of sight integrations of $p_{th}(B \sin
\theta)^{3/2}$, where $\theta$ is the angle between the line of sight
and the magnetic field (Clarke 1989).  Such an approximation is
necessary when relativistic particles are not tracked explicitly.  The
total intensity is integrated along the $y$-direction and is plotted on
the $xz$-plane in the figure.  In order to compare the morphology of
the images, we have used different scales in each panel of Figure
\ref{radio}.  In each panel, the range covers 3 orders of magnitude and
the grayscale maximum is 20\% above the actual maximum intensity.  The
total intensity is overlayed by B-field polarization vectors.  We note
that B-vectors are aligned with filaments in the intensity image, as
was also the case in the simulations in RHCJ.  Of the three jets, the
dense equipartition jet (simulation B) has the most obvious
spine-sheath morphology in the intensity images and also maintains the
spine farthest across the grid.  The light jet simulation (simulation
C) has a more uniform intensity for $z/R_0 \gtrsim$ 40 than the dense
expanding jet simulations, which is evidence for more uniform mixing
and is consistent with a saturation of the KH instability.

In the cylindrical jets studied in  RHCJ, the slow, fast, and
saturation stages of mass entrainment and the KH instability are
associated with slow, fast, and zero expansion rates of the combined
spine and sheath in the simulated intensity image.  The transition
points between the different expansion rates, as determined from Figure
\ref{radio}, are listed for both the expanding jets and their
cylindrical equivalents in Table \ref{siglam}.  In the dense expanding
jet simulations presented here (simulations A and B), a low intensity
quickly expanding sheath appears at $z/R_0 \approx$ 20.  This rapid
expansion ceases for $z/R_0 \gtrsim$ 50 in the simulation with the
stronger field (simulation B), another indication of somewhat greater
stability.  The intensity image in simulation B appears to enter the
third (saturation) stage although no equivalent stage for the
magnetized mass ($\sigma$) appears on the grid (see Fig.\ \ref{sig}).
This suggests that this simulation continues to entrain mass for some
distance beyond where the apparent expansion stops.  This
characteristic was seen in some of the cylindrical simulations of RHCJ,
although this usually occurred for light cylindrical jets.  In
simulations A and B, the transition points of the total intensity
expansion listed in Table \ref{siglam} are similar to those listed for
the linear mass density in Table \ref{tran}.  We note that the
fluctuations in the linear mass density may be associated with growing
vortices at the jet/external medium interface, and there is evidence
for such vortices in the wispy edges of the total intensity images in
simulations A and B.

There are three stages of expansion in the total intensity image from
the light jet simulation C; particularly notable is the nearly constant
width for $z/R_0 \gtrsim$ 45.  Again we see a correspondence between
the transition positions determined from the linear mass density and
from the total intensity width.  As with the transition points
demarking different mass entrainment rates, the distance between the
two transition points determined from the intensity images for the
expanding jets is larger than the distance between the transition
points for the cylindrical jets.  This result is consistent with the
initially shorter e-folding lengths but longer distances to 5
e-foldings on the expanding jet when compared to the cylindrical jet
(see Table \ref{grow}).  Thus, jet expansion has served to partially
stabilize the lower magnetosonic Mach number expanding jet relative to
a cylindrical jet and the total intensity images show observational
consequences of this stabilization.

\subsection{Jet Structure Related to Mass Entrainment and KH Instability}

In this subsection, we compare wavelengths and wave speeds measured in
the simulations with those expected from a linear analysis of KH
induced modes.  Subsequently, we examine the effect of the KH
instability on the fluting of the jet surface via axial velocity
cross-sections and on internal structure via axial cuts of each
velocity component.

\subsubsection{Wavelengths and Wave Speed Estimates}

From the oscillations observed in the intensity images
(Fig.\ \ref{radio}), we estimate the wavelength of the surface helical
mode, $\lambda_h$, to be: in simulation A, 15$R_0$; in B, 13$R_0$; and
in C, 18$R_0$.  The error in this estimate is roughly 10\%.  We
measured these wavelengths in a region centered on $z/R_0$ = 25 in
simulations A and B, and $z/R_0$ = 20 in simulation C; recall that the
wavelength at maximum growth should be proportional to the jet radius.
The expanding jets have a smaller precessional frequency, a lower jet
velocity and a different wave speed than their cylindrical
equivalents.  These factors combine to give similar wavelengths in the
expanding jets and their cylindrical equivalents (see Table
\ref{siglam}).

If the observed wavelength of a mode is much larger than the wavelength
associated with peak growth, $\lambda_h > 10\lambda^*$, then the wave
speed should be roughly $\eta u/(1+\eta)$.  However, if $\lambda^* \le
\lambda_h \le 3\lambda^*$, the wave speed approximations for peak
growth are more appropriate, and the wave speed should be $\approx
\eta^{1/2}u/(1+\eta^{1/2})$ (e.g., Hardee 1987, and HCR).  From the
approximation (4b) in HCR for the wavelength at maximum growth, we
estimate in simulations A and B at $z/R_0$ = 30 that $\lambda^* =
10R_0$ and in simulation C at the same position $\lambda^* = 16R_0$.
Thus, the measured wavelengths indicate that the wave speed in these
simulations should be $\approx \eta^{1/2}u/(1+\eta^{1/2})$.  Assuming
that the initial values for the external density and the jet densities
follow from the expected expansion, this approximation yields the wave
speeds given in Table \ref{wavesp}.  These wave speeds are consistent
with wave speeds estimated from the product of the observed helical
mode wavelength and the driving frequency, $\nu \equiv \omega/2\pi$
(also in Table \ref{wavesp}).

In order to calculate the wave speed directly, we measure the movement
over time of maxima in axial profiles of the simulated total intensity
images (Figure \ref{intpro}).  The linear analysis approximation for
wave speed suggests that the wave speed should increase only moderately
with $z$ in the dense jet simulations.   On the other hand, the wave
speed should increase more dramatically in the light jet simulation.
Specifically, in the weak magnetic field dense jet (simulation A), the
wave speed increases from $\sim 2.3a_{ex}$ at $z/R_0 \sim$ 15 to $\sim
2.7a_{ex}$ in the 25 $\lesssim z/R_0 \lesssim$ 40 interval.  In the
more strongly magnetized dense jet (simulation B), the wave speed
increases from $\sim 2.7a_{ex}$ at $z/R_0 \sim 10$ to $\sim 2.9a_{ex}$
for 30 $\lesssim z/R_0 \lesssim$ 50.  Thus, the wave speed in the dense
jets increases by roughly 10--20\% over this range of axial positions.
In contrast, in the light jet simulation the wave speed increases from
$\sim 3.0a_{ex}$ at $z/R_0 \sim$ 15 to $\sim 7.6a_{ex}$ at $z/R_0 \sim$
30.  The large wave speed, which occurs where the KH instability has
reached nonlinear proportions, does not increase much beyond this axial
position.  It is likely that this constant wave speed is related to
saturation of the KH instability (as inferred from the saturation in
the linear mass density and the constant apparent width of the
intensity image) for $z/R_0 \gtrsim$ 40.

The measured wavelengths and wave speeds, except where the amplitude of
the KH instability has grown beyond the linear approximation, are
consistent with those estimated from the linear analysis of the KH
instability.  Thus, we conclude that the KH instability is responsible
for the growth of mass entrainment in these simulations.

\subsubsection{Cross-Sections of Axial Velocity}

The growth of the KH instability and its effect on the jet is
dramatically demonstrated in the grayscale cross-sections of axial
velocity from simulations A, B, and C in Figure \ref{cross}.  As
observed in magnetic pressure cross-sections in previous simulations
(RHCJ), there are many corrugations in the fast moving surface at
$z/R_0 \simeq$ 15 in all three simulations.  That the many corrugations
are roughly evenly spaced circumferentially on the jet is evidence for
the surface waves of high order fluting modes.   In all the simulations
here and in simulations in RHCJ, the higher order modes that dominate
close to the inlet are overwhelmed by the larger distortion amplitudes
accompanying the slower growing, lower order modes farther down the
jet.  In Figure \ref{cross} the helical mode appears as the clockwise
motion of the jet center about the initial jet axis as one moves down
the jet.   This clockwise rotation about the jet axis provides an
additional estimate for the wavelength of the helical mode.   In the
light jet simulation, $\lambda_h/R_0 \approx 35$ between
$z/R_0 = 25$, where much of the jet is to the left of the original jet
axis, and $z/R_0 = 40$, where the jet has been displaced to the upper
right.  This wavelength is consistent with the very large wave speed
($v_w/a_{ex}$ = 7.6) measured at large $z/R_0$.

There is also a correspondence between features in the intensity images
and the axial velocity cross-sections.  Recall that the simulated
intensity images are integrated along the $y$ axis and the observer in Figure
\ref{radio} is to the right in Figure \ref{cross}.   
For the light expanding jet simulation, a spine-sheath
morphology is more evident in the cross-sections than in the intensity images.   However, far from the inlet the area of the high $v_z$ material
is smaller than for the dense expanding jets.  This is another
indication of greater mixing occurring in the light jet simulation.

\subsubsection{1D Axial Cuts of Velocity}

The effect of the KH instability on the internal structure of the jets
is demonstrated by the one-dimensional cuts of velocity components
along the jet axis ($x, y$ = 0) shown in Figure \ref{axis}.  In all
three simulations, the variation in the axial velocity component is
initially small, $\lesssim$ 2\%.  At some point the variation grows to
significant amplitude relatively abruptly.  For the dense jet
simulations this occurs at $z/R_0 \sim$ 22, and for the light jet
simulation this occurs at $z/R_0 \sim$ 15.  These positions are close
to the first transition point (in either magnetized mass or width of
the simulated intensity image) in all three jets.   Thus, changes
in the nature of the mass entrainment rate associated with the growth
of the KH instability are able to affect the jet dynamics
significantly.

Continued growth of the KH instability as illustrated by axial profiles
of $v_z$ differs in the dense jet simulations beyond the first
transition point.  Specifically, there are significant oscillations of
$v_z$ in simulation A for 20 $\lesssim z/R_0 \lesssim$ 42, but
oscillations in simulation B remain relatively small out to $z/R_0
\sim$ 47.  The smaller variations in the more strongly magnetized jet
imply greater stability.  In the light jet simulation, there is an
additional, extremely abrupt (0.75$u$) decrease in $v_z$ at $z/R_0
\sim$ 31.  Examination of the  axial velocity cross-sections in Figure
\ref{cross} reveals that this extremely abrupt drop in axial velocity
in the light jet simulation is created as the small high velocity spine
is displaced off the axis by the helical mode of the KH instability,
and is not associated with the onset of the plateau in the jet plus
entrained mass.  The second transition position in the light jet does
coincide with the reappearance of the $v_z$ maxima in Figure
\ref{axis}.

In the dense jet simulations the transverse velocity components show a
noticeable relative phase shift in the region 10 $\lesssim z/R_0
\lesssim$ 30.  We consider the maximum in $v_y$ at $z/R_0 =$ 10 in
simulation A to be a precursor of, and therefore connected to, the
maxima in $v_x$ at $z/R_0 =$ 21 and in $v_y$ at $z/R_0 \sim$ 23.  The
phase shift is more easily seen in the more strongly magnetized jet
(simulation B), with extrema in $v_x$ at $z/R_0 =$ 12, 18, and 23
repeated in $v_y$ at $z/R_0 =$ 13, 21, and 25.  In the light jet
simulation, the evidence for a relative phase shift between the
transverse velocity components is less obvious, but perhaps the maximum
in $v_x$ at $z/R_0 =$ 7 contributes to the maximum in $v_y$ at $z/R_0
=$ 10.  In all of the simulations, the extent of the region where the
phase shift is noticeable is related to the growth of the helical mode
of the KH instability: the region begins where the fluctuations in
transverse velocity reach a significant amplitude and ends where the
displacement from the KH instability has become significant.   Where
noticeable, $v_y$ is shifted farther down the jet than $v_x$.   Such a
sequence is in the sense of a clockwise spatial variation when viewed
toward +$z$ from the inlet and consistent with the counterclockwise
temporal precession at the inlet described in \S 2.  This phase shift
is additional evidence for a significant presence of the helical KH
mode.

\section{CONCLUSIONS}

The magnetized, or jet plus entrained, mass in these expanding jet
simulations increases more slowly than in their cylindrical
equivalents.  This reduction in the mass entrainment is a direct
consequence of the reduced external mass density, which is required by
an expanding pressure-matched jet.  For the parameters that we have
simulated, the reduction of magnetized mass in the expanding jet
simulations appears similar to (for the dense jets) or greater than
(for the light jet) the reduction caused by changing the magnetic field
configuration from primarily axial to primarily toroidal in the
cylindrical jet simulations of RHCJ.  Thus, a conical jet appears more
stable than its cylindrical equivalent, and should be considered,
along with relativistic jets (Hardee et al.\ 1998; and Hughes, Miller,
\& Duncan 1999), as another stabilizing influence in the propagation of
astrophysical jets.

As in the cylindrical jet simulations, the spatial development of the
magnetized mass passes through different stages, in which the spatial
rate of increase of magnetized mass or the smoothness of this increase
is different.  We see evidence for a first transition, between a region
of a nearly constant increase in magnetized mass and a more varied
increase in magnetized mass in the dense jets.  In the light jet
simulation this first transition occurs between a region with a slow
increase in magnetized mass to one with a faster increase.  Also, in
the light jet simulation we see a second  transition, between the
region of fast increase and a region where the magnetized mass remains
relatively constant.  In the expanding jet simulations the distance
between first and second transition points is similar to or larger than
this distance in the cylindrical jet equivalent simulations.  The
observed behavior is consistent with slower growth of the KH
instability on expanding jets relative to cylindrical jets.  Thus, we
confirm the theoretical prediction that the KH instability grows on
these expanding jets at a reduced rate relative to an equivalent
cylindrical jet.  Previous work (RHCJ) associated the different stages
of spatial development with the linear, nonlinear, and saturated stages of
the KH instability, and this appears to be the case in these expanding
jet simulations.  The simulations also contain wavelengths and wave
speeds that are consistent with theoretical estimates of the
wavelengths and wave speeds appropriate to the surface helical mode 
wave triggered at the precession frequency. 

Simulated intensity images also reveal three stages in the expansion of
the mixed jet and entrained material.  These stages are similar (slow,
fast, and zero expansion) and roughly coincident with the three stages
of mass entrainment.  However, in one case the jet continues to entrain
mass while maintaining a constant jet width in the intensity image.
The reduced KH instability growth rate in expanding jets does result in
a lengthening span between transition points as determined from the
intensity image relative to transition points for cylindrical
equivalent jets.

Our present study confirms results from RHCJ.  The simulated total
radio intensity images and the axial velocity transverse
cross-sections show that dense jets are able to maintain a high
velocity spine as part of a spine/sheath structure.  We also see an
alignment of polarization vectors with filaments in the simulated
intensity images.  In addition, the transverse cross-sections show a
progression from high order, fast growing, small amplitude to low
order, slower growing, large amplitude KH modes.  Axial cuts of each
velocity component show that the internal structure exhibits a
transition in the growth of the instability in roughly the same
position as the first transition point in the mass entrainment rate or
in the width of the intensity images.  Beyond this transition point,
especially in the light jet simulation, the displacement of the jet
from the initial axis complicates the profile.

\acknowledgements {A.\ Rosen and P.\ Hardee acknowledge support from
the National Science Foundation through grant AST-9802955 to the
University of Alabama. The authors would also like to acknowledge David
Clarke who has provided valuable support through development and
maintenance of ZEUS-3D.  The numerical work utilized the Cray C90 at
the Pittsburgh Supercomputing Center and the storage facilities
associated with the Cray T90 at the San Diego Computing Center.}



\figcaption{The normalized linear mass density $\sigma$ (top and middle
panels) and average axial velocity $\langle v_z\rangle$ (bottom
panel).  Also shown is $\sigma$ from the cylindrical equivalent
simulations in RHCJ.  In order to show the extent of the second (fast
growing) stage in the light jet simulations, estimates of the mass
entrainment rates have been added in this stage to the middle panel.
\label{sig}}

\figcaption{The average density of the entrained mass, $\rho_{en}$, in
the expanding jet simulations normalized by the entrained mass density in 
each equivalent cylindrical jet simulation.  The solid line is the initial
profile of the external density, the dotted line is the calculated mean
density of magnetized mass less the jet mass. 
\label{aved}}

\figcaption{Integrated simulated intensity images overlayed with
B-field polarization vectors. The total intensity is shown in
grayscale, and covers three orders of magnitude with a maximum set to
roughly 1.20 times the actual maximum intensity in the image.  In
addition, the polarization vectors have a length proportional to the
fractional polarization and are shown only where the total intensity is
above 0.001 of the scaled maximum intensity in the image.
\label{radio}}

\figcaption{Evolution of total intensity profiles on the $z$-axis.  The
simulated total intensity, which is given in a quasi-logarithmic scale,
is displayed for three different times for each simulation.
\label{intpro}}

\figcaption{Grayscale cross-sections of axial velocity in the
$xy$-plane for simulation A (top), B (middle), and C (bottom).  Here,
the $+x$ axis is towards the top of each panel and
$+y$ to the right ($+z$ is into the page), and each axis spans the
region -4 $< x/R_0, y/R_0 <$ 4.  Darker shading indicates larger $v_z$,
white indicates $v_z \le$ 0.  The number in the
upper left of each panel indicates the axial position, $z/R_0$.
\label{cross}}

\figcaption{One dimensional cut of each velocity component along the
$z$-axis.  The panels on the left show the axial profile of $v_z$ and
the panels on the right show $v_x$ and $v_y$, all in units of the external medium sound speed, $a_{ex}$.  Note the different ranges of velocity
on each panel, although this range is approximately 0.25$u$ in all of
the panels.
\label{axis}}


\begin{deluxetable}{ccccccccccc}
\tablewidth{0pt}
\tablecaption{ Inlet Jet Parameters \label{para}  }
\tablehead{ 
\colhead{Sim.} 
& \colhead{($\eta$)\tablenotemark{a}} 
& \colhead{({\bf B})\tablenotemark{b}} 
& $(p_{jt}/p_{ex})$\tablenotemark{c} 
& \colhead{$(a_{jt}/a_{ex}$)\tablenotemark{c}} 
& \colhead{($V_{A}/a_{ex}$)\tablenotemark{d}} 
& \colhead{($a^{ms}_{jt}/a_{\rm ex}$)\tablenotemark{c,d}} 
& \colhead{$M_{ex}$} 
& \colhead{($M_{jt}$)\tablenotemark{c}} 
& \colhead{($M_{A})$\tablenotemark{d}} 
& \colhead{($M^{ms}_{jt})$\tablenotemark{c,d}} \nl
}
\startdata
A & 4.00   & W & 0.995 & 0.50 & 0.04 & 0.50 & ~3.49 & ~7.00 & 93.1~ & 6.98 \nl
B & 4.00   & E & 0.54~ & 0.37 & 0.38 & 0.52 & ~3.67 & 10.00 & ~9.80 & 7.00 \nl
C & 0.25   & E & 0.54~ & 1.47 & 1.50 & 2.10 & 11.17 & ~7.60 & ~7.45 & 5.32 \nl
\enddata
\tablenotetext{a} {$\eta \equiv \rho_{jt}/\rho_{ex}$, uniform jet density
across the jet.}
\tablenotetext{b} {Descriptions of the magnetic field: E=Equipartition ($B_z = 0.75, B^{pk}_{\phi} = 0.26$, in ZEUS-3D units), W=Weak ($B_z = 0.075, B^{pk}_{\phi} = 0.035$, in ZEUS-3D units).}
\tablenotetext{c} {Jet thermal pressure used is before slight modification 
(based on Equation [1]) to a uniform profile across jet.}
\tablenotetext{d} {Based on axial magnetic field only, valid for $r$ = 0 and 0.9 $\le r/R_0 \le$ 1.0.}
\end{deluxetable}


\begin{deluxetable}{cccccc}
\tablewidth{0pt}
\tablecaption{Precessional Data \label{prec} }
\tablehead{
\colhead{Simulation}  
& \colhead{$\omega R/a_{ex}$} 
& \colhead{ ($\omega R/a_{ex})_{cyl}$\tablenotemark{a} } 
& \colhead{$\tau_{p}$} 
& \colhead{$(v_{t}/a_{ex})$\tablenotemark{b} } 
& \colhead{$(v_{t}/a^{ms}_{jt,0})$\tablenotemark{b} }\nl 
}
\startdata 
A &  0.9 & 1.0 & 6.98  & 0.019 & 0.039   \nl
B &  1.1 & 1.2 & 5.71  & 0.020 & 0.039   \nl
C &  1.4 & 1.6 & 4.49  & 0.062 & 0.030   \nl
\enddata
\tablenotetext{a}{Frequencies used in previous cylindrical jet simulations (RHCJ).}
\tablenotetext{b}{Only the transverse velocity related to 
the precession, does not include velocity associated with the jet 
expansion.}
\end{deluxetable}

\begin{deluxetable}{ccccc}
\tablewidth{0pt}
\tablecaption{Magnetized Mass Related Quantities \label{tran} }
\tablehead{
& \multicolumn{2}{c}{$\sigma(z/R_0 = 60)$}  & \multicolumn{2}{c}{Transition
Points} \nl 
\colhead{Simulation} & \colhead{Conical} & \colhead{Cylindrical\tablenotemark{a}} & \colhead{Conical} & \colhead{Cyl.\tablenotemark{a}} \nl 
}
\startdata
A  &  1.75--2.0 &  2.8/3.6\tablenotemark{b}&  20--25$^c$/$>$60  &  25--27/48 \nl
B  &  1.5       &  2.8                     &  20--25$^c$/$>$60  &  25/$>$60  \nl
C  &  7--8      &  25                      &  10/42    &  18/40     \nl
\enddata
\tablenotetext{a}{Cylindrical jet data from RHCJ.}
\tablenotetext{b}{The two values listed here depend on whether the
jet reached saturation at $z/R$ = 60 (2.8) or not (3.6). Both
interpretations are possible.}
\tablenotetext{c}{Based on position where ever increasing 
fluctuations in $\sigma$ begin.  Note that similar fluctuations occur
at $z/R_0 \sim$ 20 as well in cylindrical jet simulations. } 
\end{deluxetable}

\begin{deluxetable}{ccccc}
\tablewidth{0pt}
\tablecaption{ Effective Growth Lengths \label{grow} }
\tablehead{
 & \multicolumn{2}{c}{Expanding Jet}  & \multicolumn{2}{c}{Cylindrical Jet} \nl 
\colhead{Simulation} & \colhead{$N_e = 1$} & \colhead{$N_e = 5$} & \colhead{$N_e = 1$} & \colhead{$N_e = 5$\tablenotemark{a}}\nl 
}
\startdata
A &  6.5 & 39.5 & 6.9 & 34.5 \nl
B &  6.5 & 40.8 & 6.9 & 34.5 \nl
C &  4.9 & 28.9 & 5.3 & 26.5 \nl
\enddata
\tablenotetext{a}{Values here are 5 times those in previous
column.}
\end{deluxetable}

\begin{deluxetable}{ccccc}
\tablewidth{0pt}
\tablecaption{Data from Simulated Intensity Maps\label{siglam} }
\tablehead{
& \multicolumn{2}{c}{Transition Points}  & \multicolumn{2}{c}{$\lambda_{h}$} \nl \colhead{Simulation} 
& \colhead{Conical} 
& \colhead{Cyl.\ \tablenotemark{a}} 
& \colhead{Conical} 
& \colhead{Cyl.\ \tablenotemark{a}}\nl 
}
\startdata
A & 20/$>$60  &  25/45        &  15 &  13   \nl
B & 20/50    &  30/50        &  13 &  13   \nl
C & 15/45    &  $\sim$20/40  &  18 &  20   \nl
\enddata
\tablenotetext{a}{Cylindrical data taken from equivalent simulations in
RHCJ.} 
\end{deluxetable}

\begin{deluxetable}{cccc}
\tablewidth{0pt}
\tablecaption{Estimated Wave Speeds\label{wavesp} }
\tablehead{
\colhead{Simulation} 
& \colhead{$v_w(z/R_0 = 0)/a_{ex}$} 
& \colhead{$v_w(z/R_0 = 20)/a_{ex}$} 
& \colhead{$\lambda_h \omega/(2\pi a_{ex})$}
}
\startdata
A & 2.3 & 2.4 &  2.2  \nl
B & 2.5 & 2.6 &  2.3  \nl
C & 3.7 & 4.0 &  4.0  \nl
\enddata
\end{deluxetable}


\begin{references}

\reference{} Appl, S., \& Camenzind, M. 1992, \aap, 256, 354

\reference{} Bassett, G.M., \& Woodward, P.R. 1995, \apj, 441, 582

\reference{} Bicknell, G.V. 1994, \apj, 422, 542

\reference{} ---------. 1996, in ASP Conf.\ Ser.\ 100: Energy Transport 
in Radio Galaxies and Quasars, eds.\ P.E.\ Hardee, A.H.\ Bridle, \& J.A. Zensus, (San Francisco:ASP) 253

\reference{} Biretta, J.A., Zhou, F., \& Owen, F.N. 1995, \apj, 447, 582 

\reference{} Bodo, G., Massaglia, S., Rossi, P., Rosner, R., Malagoli,
A., \& Ferrari, A. 1995, \aap, 303, 281

\reference{} Bodo, G., Rossi, P., Massaglia, S., Ferrari, A., Malagoli, A.,
\& Rosner, R. 1998, \aap, 333, 1117

\reference{} Clarke, D.A. 1989, \apj, 342, 700

\reference{} ---------. 1996, \apj, 457, 291

\reference{} DeYoung, D.S. 1996, in ASP Conf.\ Ser.\ 100: Energy Transport
   in Radio Galaxies and Quasars, eds. P.E. Hardee, A.H. Bridle, \& 
   J.A. Zensus, (San Francisco: ASP) 261

\reference{} Fanaroff, B.L., \& Riley, J.M. 1974, \mnras, 167, 31P

\reference{} Giovannini, G., Taylor, G.B., Arbizanni, E., Bondi, M., Cotton, W.D., Feretti, L., Lara, L., \& Venturi, T. 1999, \apj, 522, 101

\reference{} Hardee, P.E. 1986, \apj, 303, 111 

\reference{} ---------. 1987, \apj, 313, 607 

\reference{} Hardee, P.E., Clarke, D.A., \& Rosen, A. 1997, \apj, 485, 533 (HCR)

\reference{} Hardee, P.E., \& Rosen, A. 1999, ApJ, 524, 650

\reference{} Hardee, P.E., Rosen, A., Hughes, P.A., Duncan, G.C.  1998, \apj, 500, 598

\reference{} Hughes, P.A., Miller, M.A., \& Duncan, G.C.  1999, \baas, 31, 1547

\reference{} Katz-Stone, D.M., Rudnick, L., Butenoff, C., \& O'Donoghue, A.A. 1999, \apj, 516, 716

\reference{} Loken, C., Burns, J.O., Bryan, G., \& Norman, M. 1996 in
   ASP Conf.\ Ser.\ 100: Energy Transport in Radio Galaxies and Quasars, eds.
   P.E. Hardee, A.H. Bridle, \& J.A. Zensus, (San Francisco:ASP) 267


\reference{} Perley, R.A., Bridle, A.H., \& Willis, A.G. 1984, \apjs, 54, 291


\reference{} Rosen, A., Hardee, P.E., Clarke, D.A., \& Johnson, A.  1999, \apj, 510, 136 (RHCJ)



\reference{} Stone, J.M., Hawley, J.F., Evans, C.E., \& Norman, M.L. 1992, \apj, 388, 19

\reference{} Ulvestad, J.S., Wrobel, J.M., Roy, A.L., Wilson, A.S., Falcke, H.,
\& Krichbaum, T.P. 1999, \apj, 517, L81

\reference{}  van Leer, B.\ 1977, J.\ Comput.\ Phys., 23, 276



\end{references}
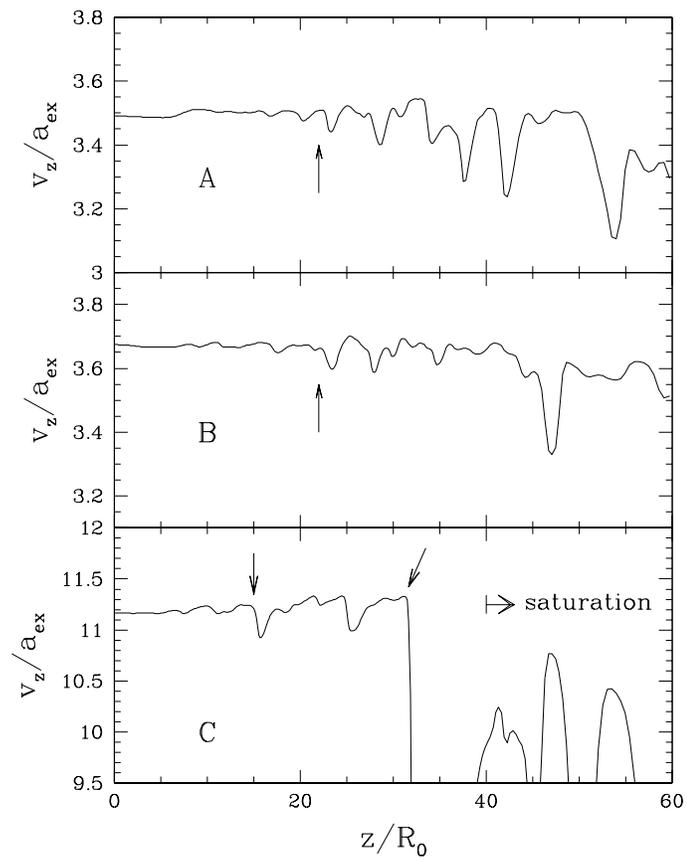
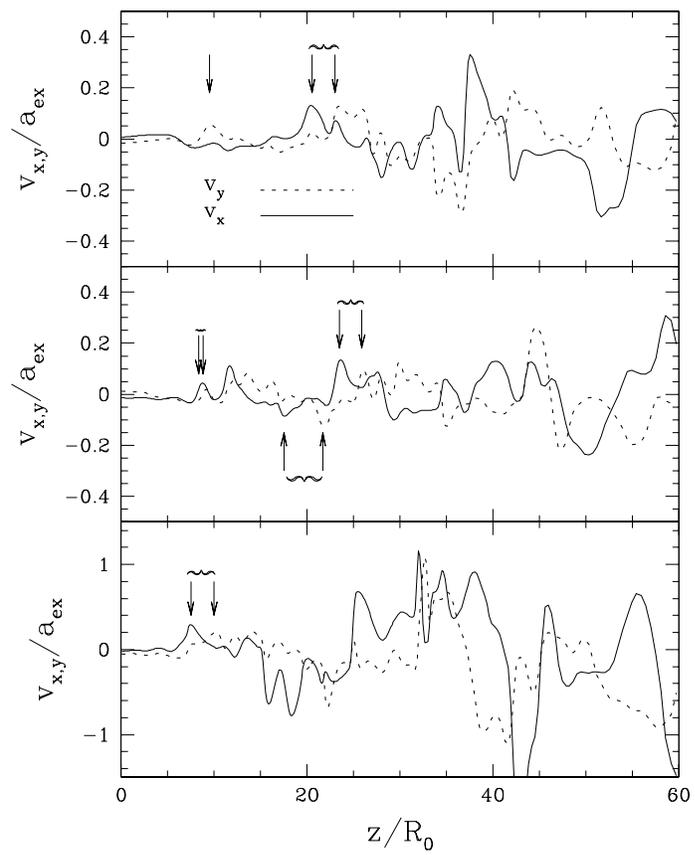
\end{document}